\documentclass[aps,nofootinbib,prd,preprint,preprintnumbers]{revtex4}

\pdfoutput=1 

\usepackage[T1]{fontenc} 
\usepackage{graphicx}
\usepackage{epsfig}
\usepackage{rotating}
\usepackage{color}
\usepackage[normalem]{ulem}
\usepackage{multirow}
\usepackage{tabularx}
\usepackage{graphicx}
\usepackage{float}
\usepackage{amsmath}
\usepackage{amssymb}
\usepackage{ulem}
\usepackage{color}

\def\SM{$\mathrm{SU(3)_c \otimes SU(2)_L \otimes U(1)_Y}$ }

\newcommand{\AddrAHEP}{
  {\it AHEP Group, Instituto de F\'{\i}sica Corpuscular --
    C.S.I.C./Universitat de Val{\`e}ncia \\
    Edificio de Institutos de Paterna,
 C/Catedratico Jos\'e Beltr\'an, 2 E-46980 Paterna (Val\`{e}ncia) - SPAIN}}

\begin{document} 

\title{Probing neutrino magnetic moments at Spallation
  Neutron Source facilities}

\author{T.S. Kosmas~$^1$}\email{hkosmas@uoi.gr}
\author{O.G. Miranda~$^2$}\email{omr@fis.cinvestav.mx}
\author{D.K. Papoulias~$^{1,3}$}\email{dimpap@cc.uoi.gr}
\author{M. T\'ortola~$^3$}\email{mariam@ific.uv.es}
\author{J.W.F. Valle~$^3$}\email{valle@ific.uv.es, URL:
  http://astroparticles.es/} 

 \affiliation{$^1$~Theoretical Physics
  Section, University of Ioannina,\\GR-45110 Ioannina, Greece}
\affiliation{$^2$~Departamento de F\'{\i}sica, Centro de Investigaci\'on
  y de Estudios Avanzados del IPN,\\ Apartado Postal 14-740 07000 Mexico,
  Distrito Federal, Mexico}
\affiliation{$^3$~\AddrAHEP} 

\begin{abstract} 
  Majorana neutrino electromagnetic properties are studied
  through neutral current coherent neutrino-nucleus scattering. We
  focus on the potential of the recently planned COHERENT experiment
  at the Spallation Neutron Source to probe muon-neutrino magnetic
  moments. The resulting sensitivities are determined on the basis of
  a $\chi^2$ analysis employing realistic nuclear structure
  calculations in the context of the quasiparticle random phase
  approximation. We find that they can improve existing limits
  by half an order of magnitude. In addition, we show that these
  facilities allow for standard model precision tests in the low
  energy regime, with a competitive determination of the weak mixing
  angle. Finally, they also offer the capability to probe other
  electromagnetic neutrino properties, such as the neutrino
  charge radius. We illustrate our results for various choices of
  experimental setup and target material.  
\end{abstract}


\maketitle

\section{Introduction}
\label{sec:intro}

The robust confirmation of the existence of neutrino masses and
mixing~\cite{Schechter:1980gr,Schechter:1981cv}, thanks to the
milestone discovery of neutrino oscillations in propagation from
solar, atmospheric, accelerator, and reactor neutrino sources, has
opened a window to probe new physics beyond the standard model
(SM)~(for the relevant experimental references see e.g.
\cite{Maltoni:2004ei,Nunokawa:2007qh}). While the ultimate origin of
neutrino mass remains a mystery~\cite{Valle:2015pba}, oscillation
results have motivated a plethora of SM extensions to generate small
neutrino masses~\cite{Boucenna:2014zba}. A generic feature of such
models is the existence of nontrivial neutrino electromagnetic (EM)
properties~\cite{Schechter:1981hw,Shrock:1982sc,kayser:1982br,Nieves:1981zt,Beacom:1999wx,Broggini:2012df}. 
Although the three-neutrino oscillation paradigm seems to be on rather
solid ground~\cite{Tortola:2012te,Forero:2014bxa},  nontrivial
neutrino electromagnetic properties may still play an important
subleading role in precision neutrino studies \cite{Beringer:1900zz}.

The lowest-order contribution of neutrino EM interactions involves
neutrino magnetic moments
(NMM)~\cite{Beacom:1999wx,Miranda:2003yh,miranda:2004nz}, as well as
the neutrino
charge radius~\cite{Hirsch:2002uv,bernabeu:2000hf,Barranco:2007ea}
arising from loop-level radiative
corrections~\cite{Bardeen:1972vi,Lee:1973fw}. Note that a direct neutrino
magnetic moment measurement could provide a key insight in the
understanding of the electroweak interactions, and the Majorana nature
of neutrinos~\cite{Grimus:2002vb,Tortola:2004vh}.  Indeed, in contrast
to the case of Majorana neutrinos, only massive Dirac neutrinos can
have non-vanishing diagonal magnetic
moments~\cite{Schechter:1981hw,kayser:1982br,Nieves:1981zt,Broggini:2012df}. In the
general Majorana case, only off-diagonal transition magnetic moments
exist, they form an antisymmetric matrix, calculable from first
principles, given the underlying gauge theory.

Within the minimally extended \SM model with Dirac neutrino
  masses, one expects tiny NMM, of the order of $\mu_{\nu} \leq
  10^{-19}\mu_{B} \left(\frac{m_{\nu}}{1 \mathrm{eV}} \right)$,
  expressed in Bohr magnetons
  $\mu_B$~\cite{Lee:1977tib,Vogel:1989iv}. However, appreciably larger
  Majorana neutrino transition magnetic moments are expected in many
  theoretical models, such as those involved with left-right
  symmetry~\cite{Czakon:1998rf}, scalar
  leptoquarks~\cite{Povarov:2007zz}, R-parity-violating
  supersymmetry~\cite{Gozdz:2012xw}, and large extra
  dimensions~\cite{Mohapatra:2004ce}.  Currently, the most stringent
  upper limits, $\mu_{\nu}\leq$~few~$\times10^{-12}\mu_{B}$, come from
  astrophysics~\cite{Raffelt:1999gv,PhysRevLett.111.231301,Miranda:2003yh,miranda:2004nz}.
In addition, there are bounds from measurements by various terrestrial
neutrino scattering experiments. The present status of such
constraints is summarized in Table~\ref{table:constraints} where one
can see that the direct constraints on $\mu_{\nu_\mu}$ and
$\mu_{\nu_e}$ are still rather poor. It should be mentioned, however,
that currently operating reactor neutrino experiments such as TEXONO
and GEMMA have set robust constraints on $\mu_{\bar{\nu}_e}$.

\begin{table*}[t]
\centering
\caption{Summary of the current  90\% C.L. constraints 
  on neutrino magnetic moments from various experiments. }
\begin{tabular}{llcc}
\hline \hline
Experiment & Reaction& Observable & Constraint ($10^{-10}\mu_B$) \\
\hline
LSND~\cite{Auerbach:2001wg} & $\nu_\mu e^- \to \nu_\mu e^-$ & $\mu_{\nu_\mu}$ & $6.8$ \\
LAMPF~\cite{Allen:1992qe} & $\nu_e e^- \to \nu_e e^-$ & $\mu_{\nu_e}$ & $10.8$ \\
TEXONO~\cite{Wong:2006nx} & $\bar{\nu}_e e^- \to \bar{\nu}_e e^-$ & $\mu_{\bar{\nu}_e}$ & $0.74$ \\
GEMMA~\cite{Beda:2012zz} & $\bar{\nu}_e e^- \to \bar{\nu}_e e^-$ & $\mu_{\bar{\nu}_e}$ &$0.29$ \\
 \hline \hline
\end{tabular}
\label{table:constraints}
\end{table*}

The possibility of probing neutrino EM parameters,
  such as the NMM and the neutrino charge radius, through coherent
  elastic neutrino-nucleus scattering
  (CENNS)~\cite{Freedman:1973yd,Drukier:1983gj} can be explored on the
  basis of a sensitivity $\chi^2$-type
  analysis~\cite{Miranda:2004nb,Scholberg:2005qs,Barranco:2005yy,Barranco:2005ps,Barranco:2007tz,Escrihuela:2009up,Escrihuela:2011cf}.
  To this end here we perform realistic nuclear structure
  calculations~\cite{Kosmas:1994ti,Kosmas:2001ia} in order to compute
  accurately the relevant cross sections
  ~\cite{Papoulias:2013cba,Papoulias:2013gha,Papoulias:2015vxa}. The
  required proton and neutron nuclear form factors are reliably
  obtained within the context of the  quasiparticle random phase
  approximation (QRPA) method by considering
  realistic strong nuclear
  forces~\cite{Donnelly:1976fs,Chasioti2009234,Giannaka:2013dba,Giannaka:2015sta}.
Concentrating on ongoing and planned neutrino experiments, we have
devoted special effort in estimating the expected number of CENNS
events with high significance.  Specifically, our study is focused on
the proposed detector materials of the COHERENT
experiment~\cite{Bolozdynya:2012xv,Akimov:2013yow} at the Spallation
Neutron Source (SNS)~\cite{SNS}. Even though a CENNS event has never been experimentally measured, we remark that the highly intense neutrino
beams~\cite{Avignone:2003ep,Efremenko:2008an} provided at the SNS 
indicate very encouraging prospects towards the detection of this
reaction for the first
time~\cite{Brice:2013fwa,Collar:2014lya}, by using
low energy detectors~\cite{2013JInst...8R4001C}. Furthermore, neutrinos
from stopped pion-muon beams~\cite
{AguilarArevalo:2008yp,Louis:2009zza} at the
SNS~\cite{Amanik:2009zz,Scholberg:2009ha}
    or elsewhere~\cite{Soma:2014zgm,BNB} have
motivated many studies searching for physics beyond the SM model
too~\cite{Barranco:2007tz,Anderson:2012pn,Papoulias:2015vxa,Papoulias_to_be_published}.

In the present work we quantify the prospects, not only of
  detecting CENNS events at the SNS, but also of performing precision
  electroweak measurements and probing neutrino properties beyond the
  SM. We conclude that the extracted sensitivities on the effective
  NMM improve with respect to previous results of studies of this
  type. We obtain for the first time robust upper limits on
$\mu_{\nu_\mu}$. Moreover, we obtain a sensitivity for the
neutrino charge radius, which is competitive with those of previous
studies. Furthermore, we explore the
sensitivity of these experiments for standard model precision
measurements of the weak mixing angle in the energy regime of few MeV.


\section{Coherent elastic neutrino-nucleus scattering}

The CENNS is described
within the SM starting from the neutrino-quark neutral-current (NC)
interaction, but is expected to have corrections coming from new
physics~\cite{Maltoni:2004ei}, such as nonstandard
interactions~\cite{Miranda:2004nb,Barranco:2005yy,
Barranco:2005ps,Barranco:2007tz,
Escrihuela:2009up,Escrihuela:2011cf,Papoulias:2013gha,Papoulias:2015vxa}
or nontrivial neutrino electromagnetic properties \cite{Papoulias_to_be_published}. Here we focus on
the latter~\cite{Miranda:2003yh,miranda:2004nz,Barranco:2007ea}.

\subsection{Standard model prediction}
\label{sec:stand-model-pred}

At low and intermediate neutrino energies $E_\nu \ll M_W$, the weak
neutral-current cross section describing this process in the SM is
given by the four-fermion effective interaction Lagrangian,
$\mathcal{L}_{\mathrm{SM}}$,
\begin{equation}
\mathcal{L}_{\mathrm{SM}} = - 2 \sqrt{2} G_{F} \sum_{ \begin{subarray}{c} f= \, u,d \\ \alpha= e, \mu,\tau \\ P = L,R 
\end{subarray}} g_{\alpha \alpha}^{f,P}   
\left[ \bar{\nu}_{\alpha} \gamma_{\rho} L \nu_{\alpha} \right] \left[ \bar{f} \gamma^{\rho} P f \right] \, ,
\label{SM_Lagr}
\end{equation}
where $P=\{L,R\}$ are the chiral projectors, $\alpha=\{e,\mu,\tau \}$
denotes the neutrino flavor, $f$ is a first generation quark, and
$G_F$ is the Fermi constant. The left- and right-hand coupling
constants for the $u$- and $d$-quark to the $Z$-boson including the relevant
radiative corrections are given as~\cite{Beringer:1900zz}
\begin{equation}
\begin{aligned}
g_{\alpha \alpha}^{u,L} =& \rho_{\nu N}^{NC} \left( \frac{1}{2}-\frac{2}{3} \hat{\kappa}_{\nu N} \hat{s}^2_Z \right) + \lambda^{u,L} \, ,\\
g_{\alpha \alpha}^{d,L} =& \rho_{\nu N}^{NC} \left( -\frac{1}{2}+\frac{1}{3} \hat{\kappa}_{\nu N} \hat{s}^2_Z \right) + \lambda^{d,L} \, ,\\
g_{\alpha \alpha}^{u,R} =& \rho_{\nu N}^{NC} \left(-\frac{2}{3} \hat{\kappa}_{\nu N} \hat{s}^2_Z \right) + \lambda^{u,R} \, ,\\
g_{\alpha \alpha}^{d,R} =& \rho_{\nu N}^{NC} \left(\frac{1}{3} \hat{\kappa}_{\nu N} \hat{s}^2_Z \right) + \lambda^{d,R} \, ,
\end{aligned}
\end{equation}
with $\hat{s}^2_Z = \sin^2 \theta_W= 0.23120$, $\rho_{\nu N}^{NC} = 1.0086$, $\hat{\kappa}_{\nu N} = 0.9978$, $\lambda^{u,L} = -0.0031$, $\lambda^{d,L} = -0.0025$ and $\lambda^{d,R} =2\lambda^{u,R} = 7.5 \times 10^{-5}$.

Since neutrino detection experiments are sensitive to the kinetic
energy of the recoiling nucleus, one expresses the differential cross
section accordingly. Using the effective Lagrangian of
Eq.~(\ref{SM_Lagr}), one can describe the coherent neutrino scattering
off a spherical spin-zero nucleus of mass $M$, by computing the
differential cross section with respect to the nuclear recoil energy,
$T$, as~\cite{Papoulias:2013gha,Papoulias:2015vxa}
\begin{equation}
\left( \frac{d\sigma}{dT} \right)_{\mathrm{SM}} = \frac{G_F^2 \,M}{\pi} \left(1- 
\frac{M\,T}{2 E_\nu^2}\right)\left\vert\langle gs \vert\vert \hat{\mathcal{M}}_0(q) \vert\vert gs \rangle\right \vert^2\, , 
\label{SM_dT}
\end{equation}
where $E_\nu$ is the neutrino energy. 
Within the context of the Donnelly-Walecka multipole decomposition
method, the relevant nuclear matrix element for the dominant coherent
channel ($gs\rightarrow gs$ transitions) is based on the Coulomb
operator $\hat{\mathcal{M}}_0$~\cite{Donnelly:1976fs}. The latter is a
product of the zero-order spherical Bessel function times the
zero-order spherical harmonic~\cite{Papoulias:2013cba,Giannaka:2013dba,Giannaka:2015sta,
  Chasioti2009234} and can be cast in the form~\cite{Papoulias:2013gha}
\begin{equation}
\left\vert {\cal M}^{\mathrm{SM}}_{V} \right \vert ^{2} \, \equiv \,
\left\vert\langle gs \vert\vert \hat{\mathcal{M}}_0(q) \vert\vert gs \rangle\right \vert^2 = 
\left[g^p_V Z F_Z (q^2) + g^n_V N F_N (q^2) \right]^2 \, .
\label{SM-ME}
\end{equation} 
The finite nuclear size is taken into account by expressing the
Coulomb matrix element in terms of the proton (neutron) nuclear form
factors $F_{Z(N)}(q^2)$, reflecting the dependence of the coherent
rate on the variation of the momentum transfer, $q^2 \simeq 2 M
T$. The polar-vector couplings of protons ($g^p_V$) and neutrons
($g^n_V$) to the $Z$-boson are defined as $g^p_V = 2(g_{\alpha
  \alpha}^{u,L} + g_{\alpha \alpha}^{u,R}) + (g_{\alpha \alpha}^{d,L}
+ g_{\alpha \alpha}^{d,R})$ and $g^n_V = (g_{\alpha \alpha}^{u,L} +
g_{\alpha \alpha}^{u,R}) +2(g_{\alpha \alpha}^{d,L} + g_{\alpha
  \alpha}^{d,R})$, respectively.  It can be noticed that the vector
proton coupling, $g^p_V$, is small compared to the corresponding
neutron coupling, $g^n_V$, therefore, the dominant contribution to the
coherent cross section scales with the square of the number of
neutrons of the target isotope.

In this work we perform realistic nuclear structure calculations for
the experimentally interesting even-even nuclear isotopes, 
$^{20}$Ne, $^{40}$Ar, $^{76}$Ge and $^{132}$Xe. To this aim, the nuclear ground state, 
$\vert gs\rangle \equiv \vert 0^+\rangle$, has been constructed by solving (iteratively) the BCS
equations, quite precisely. In this framework, the proton (neutron)
nuclear form factors read~\cite{Kosmas:1994ti}
\begin{equation}
F_{N_n}(q^2) = \frac{1}{N_n}\sum_j \sqrt{2 j +1}\, \langle j\vert j_0(qr)\vert j\rangle\left(\upsilon^j_{N_n}\right)^2 \, ,
\end{equation}
where $N_{n}=Z\,\,(\mathrm{or}\,\, N)$ and $\upsilon^j_{N_{n}}$
denotes the occupation probability amplitude of the $j$th
single-nucleon level. For each nuclear system, the chosen active model
space, as well as the required monopole (pairing) residual interaction
that was obtained from a Bonn C-D two-body potential (strong
two-nucleon forces) and slightly renormalized with two parameters
$g^{p\,(n)}_{\mathrm{pair}}$ for proton (neutron) pairs, has been
taken from~Ref.~\cite{Papoulias:2015vxa}. The above method has been
successfully applied for similar calculations of various semileptonic
nuclear processes \cite{Kosmas:2001ia,Chasioti2009234,Giannaka:2015sta}.

\subsection{Electromagnetic neutrino-nucleus cross sections}

After the discovery of neutrino
oscillations~\cite{Tortola:2012te,Forero:2014bxa} over a decade ago,
it became evident that neutrinos are indeed massive
particles~\cite{Schechter:1980gr,Schechter:1981cv} and, as a result,
they may acquire nontrivial electromagnetic properties as
well~\cite{Grimus:2002vb,Tortola:2004vh}. At low-momentum transfer,
the description of possible neutrino EM interactions involves two
types of phenomenological parameters, the anomalous magnetic moment
and the mean-square
charge radius~\cite{Schechter:1981hw,Shrock:1982sc,kayser:1982br,Nieves:1981zt,Beacom:1999wx,Broggini:2012df}. It
is worth mentioning that the photon exchange involving a neutrino
magnetic moment flips the neutrino helicity, while in the interaction
due to the weak gauge boson exchange the helicity is preserved.

The electromagnetic neutrino-nucleus vertex has been comprehensively
studied~\cite{Vogel:1989iv}, and its contribution to the coherent
elastic cross section including nuclear physics details takes the
form~\cite{Papoulias_to_be_published}
\begin{equation}
\left( \frac{d \sigma}{dT} \right)_{\mathrm{EM}}=\frac{\pi \alpha_{\rm em}^2 {\mu_{eff}}^{2}\,Z^{2}}{m_{e}^{2}}\left(\frac{1-T/E_{\nu}}{T}\right) F_{Z}^{2}(q^{2})\,, 
\label{NMM-cross section}
\end{equation}
where $\alpha_{\rm em}$ is the fine structure constant and $\mu_{eff}$
is the effective neutrino magnetic moment.

In this framework, the helicity preserving standard weak interaction
cross section (SM) adds incoherently with the helicity-violating EM
cross section, so the total cross section is written as
\begin{equation}
\left( \frac{d \sigma}{dT}\right)_{\mathrm{tot}} = \left( \frac{d \sigma}{dT} \right)_{\mathrm{SM}} 
+ \left( \frac{d \sigma}{dT} 
\right)_{\mathrm{EM}}\, .
\end{equation}
The latter expression will be used below in order to
  constrain the effective neutrino magnetic moment parameters.

\section{Neutrinos from the Spallation Neutron Source}
 
There are several experimental proposals that plan to detect for the
first time a
CENNS~\cite{Amanik:2009zz,Scholberg:2009ha,Soma:2014zgm,BNB} signal.
In this section, we describe the ongoing COHERENT
experiment~\cite{Bolozdynya:2012xv,Akimov:2013yow}, proposed to
operate at the SNS at Oak Ridge National
Lab~\cite{SNS}. This facility provides excellent prospects for
measuring CENNS events for the first time. In general, any potential
deviation from the SM expectations can be directly interpreted as a 
signature of new physics and, thus, has prompted many theoretical
studies searching for physics
within~\cite{Brice:2013fwa,Collar:2014lya} and beyond the
SM~\cite{Barranco:2005yy,Scholberg:2005qs,Barranco:2007tz,Anderson:2012pn,Papoulias:2015vxa}.

\begin{table*}[b]
\centering
\caption{Summary of the detector concepts assumed in this work. We
  consider four possible nuclei as targets and two possible
  experimental setups for each nucleus, a realistic one, for
  different detector masses, distances, recoil energy windows, and
  efficiencies, and the optimistic case where all the variables are
  allowed to have their ``best'' value. }
\begin{tabular}{l|l|llll}
\hline \hline
\multicolumn{6}{ c }{COHERENT experiment} \\
\hline
& & $^{20}$Ne~\cite{Bolozdynya:2012xv} & $^{40}$Ar~\cite{Bolozdynya:2012xv} & $^{76}$Ge~\cite{Anderson:2012pn} & $^{132}$Xe~\cite{Bolozdynya:2012xv,2013JInst...8R4001C}\\ \hline
\multirow{4}{*}{Realistic} & Mass & 391 kg  & 456 kg  & 100 kg  & 100 kg  \\
  &  Distance & 46 m  & 46 m  & 20 m & 40 m  \\
 & Efficiency & 50\%  & 50\%  & 67\%  & 50\% \\
 & Recoil window & 30-160 keV  & 20-120 keV  & 10-78 keV & 8-46 keV \\ \hline
\multirow{4}{*}{Optimistic}  & mass & 1 ton & 1 ton & 1 ton & 1 ton \\
   & Distance & 20 m & 20 m & 20 m & 20 m \\
 & Efficiency & 100\% & 100\% & 100\% & 100\%\\
 & Recoil window & $1$keV -- $T_{\mathrm{max}}$ & $1$keV -- $T_{\mathrm{max}}$ & $1$keV -- $T_{\mathrm{max}}$ & $1$keV -- $T_{\mathrm{max}}$ \\ 
 \hline \hline
\end{tabular}
\label{table_consept}
\end{table*}
%

Currently, the SNS constitutes the leading facility for neutron
physics searches, producing neutrons by firing a pulsed proton beam at
a liquid mercury target~\cite{Amanik:2009zz}. In addition to
neutrons, the mercury target generates pions, which decay producing
neutrino beams as a free by-product. These beams are exceptionally
intense, of the order of $\Phi = 2.5 \times 10^{7} \, \nu \,
\mathrm{s^{-1} cm^{-2}}$ ($\Phi = 6.3 \times 10^{6} \, \nu \,
\mathrm{s^{-1} cm^{-2}}$) per flavor at 20 m (40 m) from the
spallation target~\cite{Avignone:2003ep}. In stopped
pion-muon sources, a monoenergetic muon-neutrino $\nu_{\mu}$ flux
with energy 29.9 MeV is produced via pion decay at rest $\pi^+
\rightarrow \mu^{+} \nu_{\mu} $ within $\tau=26 \, \mathrm{ns}$
(prompt flux), followed by electron neutrinos, $\nu_e$, and muon
antineutrinos, $\bar{\nu}_{\mu}$, that are emitted from the muon-decay
$\mu^{+} \rightarrow \nu_{e} e^{+} \bar{\nu}_{\mu}$ within $\tau=2.2
\, \mathrm{\mu s}$ (delayed flux)~\cite{Efremenko:2008an}. The $\nu_{e}$ and
${\bar{\nu}_{\mu}}$ neutrino spectra are described at rather high
precision by the normalized distributions~\cite{AguilarArevalo:2008yp,Louis:2009zza}
 \begin{equation}
 \begin{aligned}
\eta_{\nu_{e}}^{SNS}=& 96 E_{\nu}^{2}M_{\mu}^{-4} \left( M_{\mu}-2E_{\nu}\right)\, ,\\
\eta_{\bar{\nu}_{\mu}}^{SNS}=& 16 E_{\nu}^{2}M_{\mu}^{-4} \left( 3 M_{\mu}-4E_{\nu}\right)\, ,
\end{aligned}
\label{labor-nu}
\end{equation}
with maximum energy of $E_{\nu}^{\text{max}}=M_{\mu}/2 $,
($M_{\mu}=105.6 \, \mathrm{MeV}$ is the muon rest mass).

In this work we distinguish two cases, the optimistic and the
realistic ones. The first case is convenient for exploring the nuclear
responses of different nuclear detector isotopes, in order to get a
first idea of the relevant neutrino parameters within and beyond the
SM. The second case is useful in quantifying the sensitivities
attainable with various individual technologies of each experimental
setup. Both cases are useful and complementary to illustrate the
potential of the proposal. For instance, for the realistic case, the
original COHERENT proposal considers different detectors to be located
in different rooms and, therefore, at different distances. In
particular, the $^{132}$~Xe detector is considered to be $40$~m
from the source while other isotopes are expected to be 
$20$~m. Clearly, for shorter distances the attainable sensitivities
would be higher for any of these detectors; this possibility is
considered in the optimistic case.

In our calculations, we assume a time window of one year for the
optimistic case and $2.4 \times 10^7$ s for the realistic
case~\cite{Scholberg:2009ha}. Detailed information on the different
detector setups considered here is summarized in
Table~\ref{table_consept}. For a comprehensive description of the
relevant nuclear isotopes including the experimental criteria and
advantages of adopting each of them, the reader is referred to
Refs. \cite{Scholberg:2005qs,Papoulias:2015vxa}.


\section{Numerical results}

Assuming negligible neutrino oscillation effects in short-distance
propagation, for each interaction channel, $x=\mathrm{SM,EM,tot}$, the
total number of counts above a certain threshold, $T_{thres}$, is given
through the expression
\begin{equation}
N^{events}_{x}= K \int_{E_{\nu_\mathrm{min}}}^{E_{\nu_\mathrm{max}}} \eta^{\mathrm{SNS}}(E_{\nu})\,dE_{\nu}\int_{T_{thres}}^{T_{\mathrm{max}}} \left( \frac{d \sigma}{dT}(E_{\nu},T) \right)_{x}\, d T\, ,
\end{equation}
where $K = N_{targ} t_{tot} \Phi$, with $N_{targ}$ being the total number of
atomic targets in the detector, $t_{tot}$ the time window of data
taking, and $\Phi$ the total neutrino flux.  In the present
calculations, the various experimental concepts are taken into account
by fixing the corresponding input parameters as discussed previously.

\subsection{Standard model precision tests at SNS}

\begin{figure}[ht!]
\centering
\includegraphics[width=\textwidth]{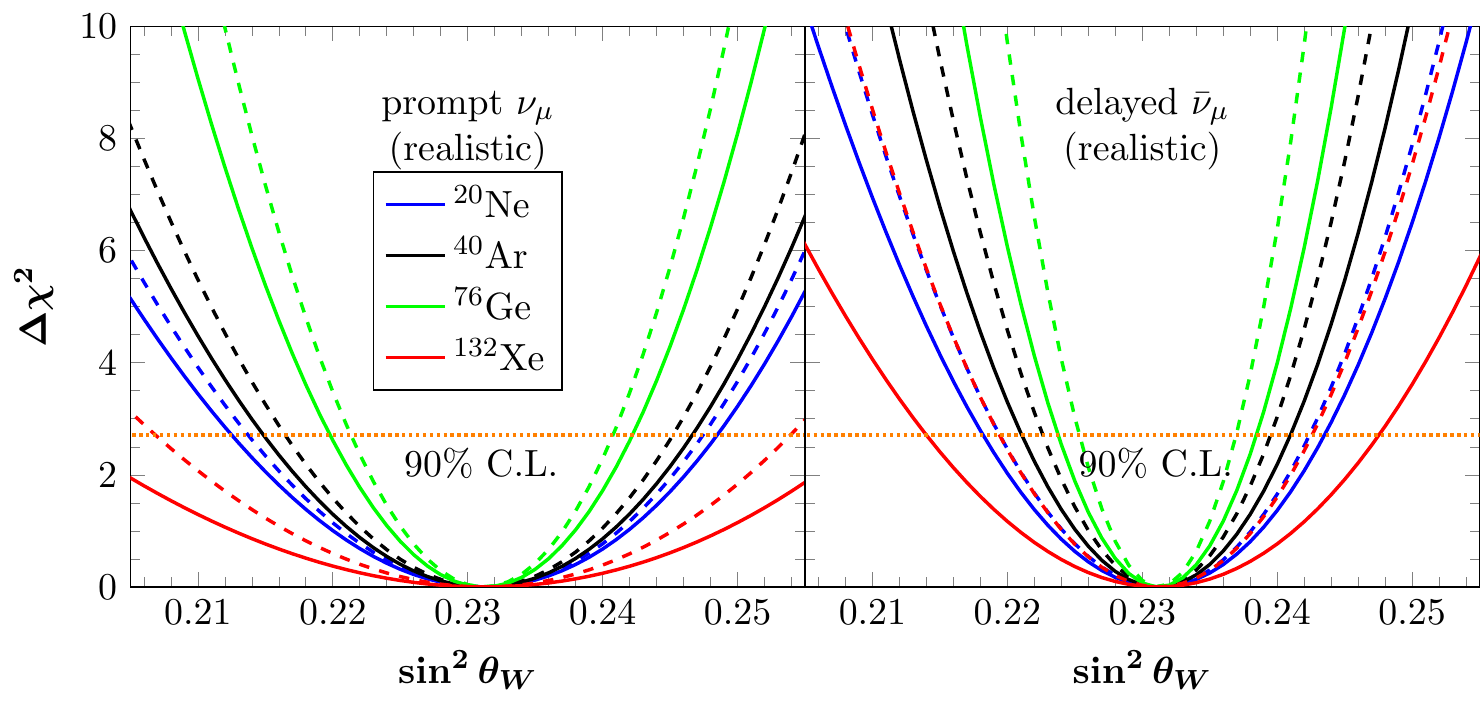}
\includegraphics[width=\textwidth]{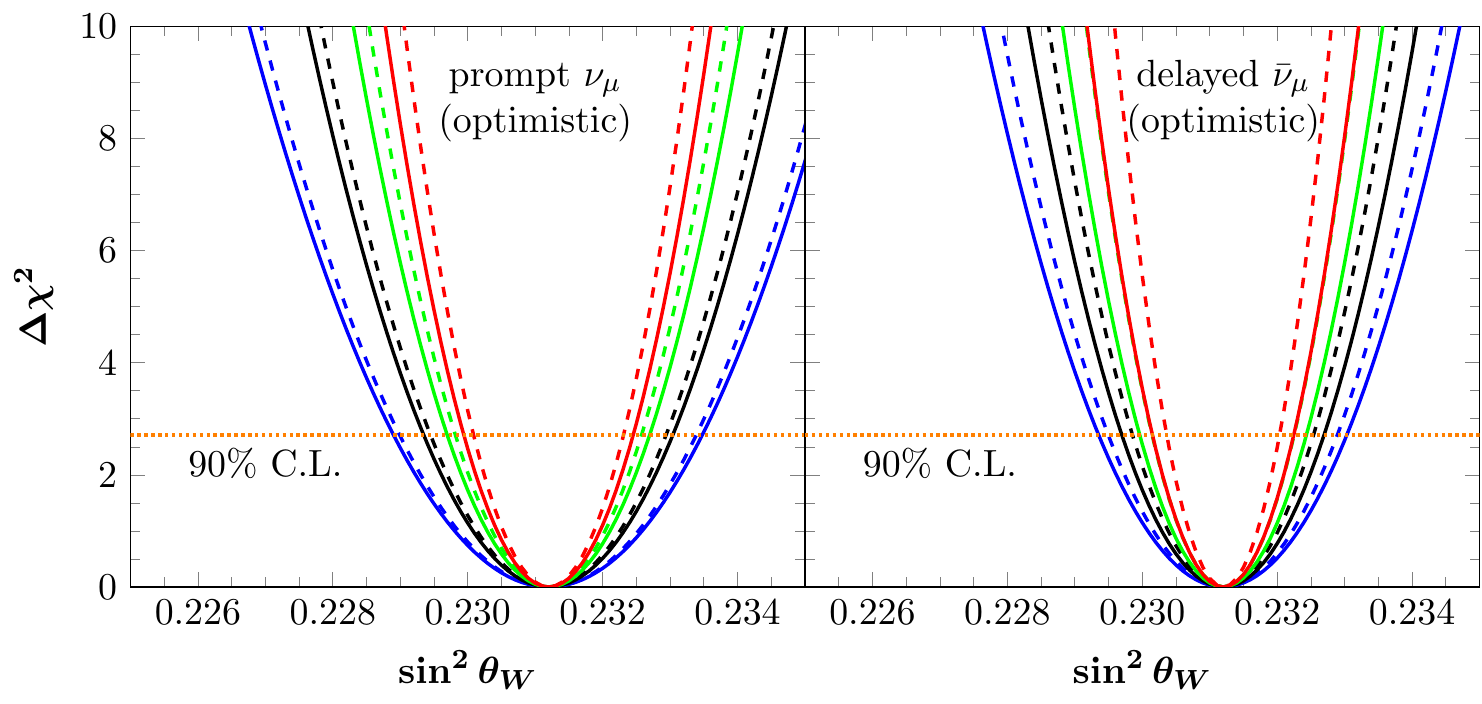}
\caption{$\Delta \chi^2$ profile in terms of the weak mixing angle
  $\sin^2 \theta_W$ showing the sensitivity of the COHERENT experiment
  to SM precision tests.  The Particle Data Group value
  $\hat{s}^2_Z=0.23120$, is used as the central value. Left (right)
  panels illustrate the results obtained by considering the prompt
  (delayed) flux, while upper (lower) panels account for the realistic
  (optimistic) case. Here, the solid (dashed) lines refer to the
  nuclear BCS method (zero-momentum transfer). }
\label{fig.sw}
\end{figure}
%
%
\begin{table*}[b]
\centering
\caption{
  Expected sensitivities to the weak mixing angle $\sin^2 \theta_W(\nu_\alpha) \equiv s^2_W(\nu_\alpha)$, assuming the various channels ($\nu_\mu,\bar{\nu}_\mu,\nu_e$) of the SNS beam for a set of possible detectors at the COHERENT experiment. For the realistic [optimistic] case, the band $ \delta s^2_W(\nu_\alpha)$ and the corresponding uncertainty are evaluated within 1$\sigma$ error. }
  \noindent
\resizebox{\linewidth}{!}{%
\begin{tabular}{c| c c| c c| c c }
\hline \hline
Nucleus & $ \delta s^2_W(\nu_\mu)$ & Uncer. (\%)   & 
          $  \delta s^2_W(\bar{\nu}_\mu)$ &  Uncer. (\%) &
          $  \delta s^2_W(\nu_e)$   &  Uncer. (\%) \\
\hline
\multirow{2}{*}{$^{20}$Ne}  
&  0.0110   &   4.74     &  0.0077    &  3.33      &  0.0091    &       3.94 \\
& [0.0014] & [0.61] & [0.0011] & [0.48] & [0.0013] & [0.56] \\   
\hline
\multirow{2}{*}{$^{40}$Ar}  
&   0.0097   &  4.17      &  0.0061    &  2.64      &   0.0074   &       3.20 \\
& [0.0011] & [0.48] & [0.0009] & [0.39] & [0.0010] & [0.43] \\ 
\hline
\multirow{2}{*}{$^{76}$Ge}   
&   0.0068   &  2.94      &  0.0045    &   1.92     &  0.0055    &       2.36 \\
& [0.0009] & [0.39] & [0.0008] & [0.35] & [0.0009] & [0.37] \\ 
\hline
\multirow{2}{*}{$^{132}$Xe} 
&    0.0181  &  7.83      &   0.0102   &   4.39     & 0.0127  &       5.47 \\
& [0.0008] & [0.35] & [0.0006] & [0.26] & [0.0007] & [0.30] \\   
 \hline \hline
\end{tabular}
}
\label{table.sw}
\end{table*}
%
%
\begin{table*}[t]
\centering
\caption{Expected sensitivities to the weak mixing angle $\sin^2 \theta_W(\nu_\mu) \equiv s^2_W(\nu_\mu)$, through a combined analysis of the prompt and  delayed beams ($\nu_\mu+\bar{\nu}_\mu$). Same conventions as in Table \ref{table.sw} are used.}
 \begin{tabular}{c| c c c c }
\hline \hline
Nucleus & $^{20}$Ne & $^{40}$Ar  &  $^{76}$Ge   &  $^{132}$Xe \\
\hline
\multirow{2}{*}{$ \delta s^2_W(\nu_\mu)$}  
&  0.0052   &   0.0042           &  0.0031    &    0.0073 \\
& [0.0007] & [0.0006] &  [0.0005] &  [0.0004] \\   
\hline
\multirow{2}{*}{Uncer. (\%)}  
& 2.23  &  1.82           &  1.34   &       3.14 \\
& [0.30] & [0.26]  & [0.22] & [0.17] \\ 
 \hline \hline
\end{tabular}
\label{table.sw2}
\end{table*}
%
%
\begin{figure}[t]
\centering
\includegraphics[width=\textwidth]{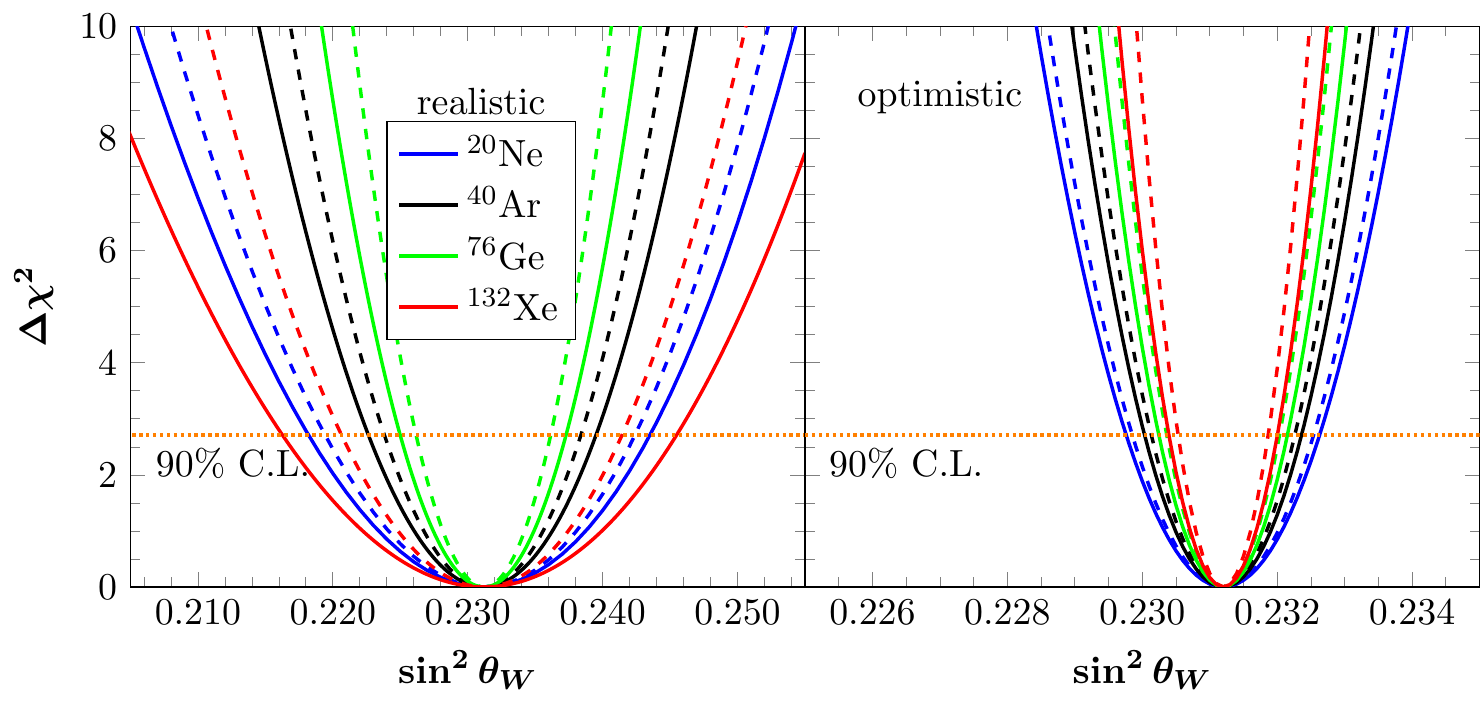}
\caption{ $\Delta\chi^2$ profile in terms of the weak mixing angle
  $\sin^2\theta_W$ from the combined measurement of the prompt and
  delayed beams ($\nu_\mu + \bar{\nu}_{\mu}$). Same conventions as in
  Fig.~\ref{fig.sw} are used.}
\label{fig.sw2}
\end{figure}
%

We first examine the sensitivity of the COHERENT experiment to the
weak mixing angle $\sin^2 \theta_W$ of the SM in the low energy regime
of the SNS operation. In order to quantify this sensitivity, assuming that the experimental proposal will measure exactly the SM
prediction, we perform a statistical analysis based on a $\chi^2$ with statistical errors only
\begin{equation}
 \chi^2 = \left( \frac{N_{SM}^{events} - N_{\text{SNS}}^{events}(\sin^2 \theta_W)}{\delta N_{SM}^{events}} \right)^2\, ,
\end{equation}
where the number of SM events, $N_{SM}^{events}$, depends on the
Coulomb nuclear matrix element entering the coherent rate. As the central
value for the SM weak mixing angle prediction we adopt the PDG value
$\hat{s}^2_Z=0.23120$. We then compute the $\chi^2$ function depending
on the expected number of events for a given value of the mixing
angle, $ N_{\text{SNS}}^{events}(\sin^2 \theta_W)$.  The corresponding
results for the various detector materials of the COHERENT experiment
are shown in Fig.~\ref{fig.sw}. In Table~\ref{table.sw}, we illustrate
the band, $\delta \sin^2 \theta_W \equiv \delta s^2_W$, at 90\%
C.L. evaluated as $\delta s^2_W=(s^2_{W^{max}}-s^2_{W^{min}})/2$ and
the corresponding uncertainty $\delta s^2_W/ \hat{s}^2_Z$, with
$s^2_{W^{max}}$ and $s^2_{W^{min}}$ being the upper and lower
$1\sigma$ error, respectively. At the optimistic level, our results
indicate that better sensitivities are expected for heavier target
nuclei, such as $^{132}$Xe.  This is understood as a direct
consequence of the significantly larger number of expected events
provided by heavier nuclear isotopes~\cite{Papoulias:2015vxa}.
However, once we consider the realistic case, the expectations change
drastically so that, for the case of a $^{76}$Ge detector we find a
better sensitivity, due to a closer location to the SNS source ($20$~m
in comparison with the $40$~m for the $^{132}$Xe case) and a higher
efficiency in recoil acceptance (see Table
\ref{table_consept}). Furthermore, in Fig. \ref{fig.sw2} and Table
\ref{table.sw2}, we show that the expected sensitivities improve
through a combined measurement of the prompt and delayed beams
($\nu_\mu + \bar{\nu}_\mu$).

\subsection{EM neutrino interactions at SNS}

One of the main goals of our present work is to examine the
sensitivity of the COHERENT experiment to the possible detection of
CENNS events due to neutrino EM effects, associated with various
effective transition NMM parameters such as
$\mu_{\nu_{\mu}}, \mu_{\bar{\nu}_{\mu}}$, and $\mu_{\nu_{e}}$. 
The total number of events expected in an experiment searching for
CENNS depends strongly on the energy
threshold $T_{thres}$ as well as the total mass of the detector. For
low energy thresholds and more massive detectors, the total number of
events expected is significantly larger and, therefore, the attainable
constraints are more stringent. We remind the reader that, for a possible NMM
detection, a very low energy threshold is required, since the EM cross
section dominates at low energies.

\begin{figure}[ht!]
\centering
\includegraphics[width=\textwidth]{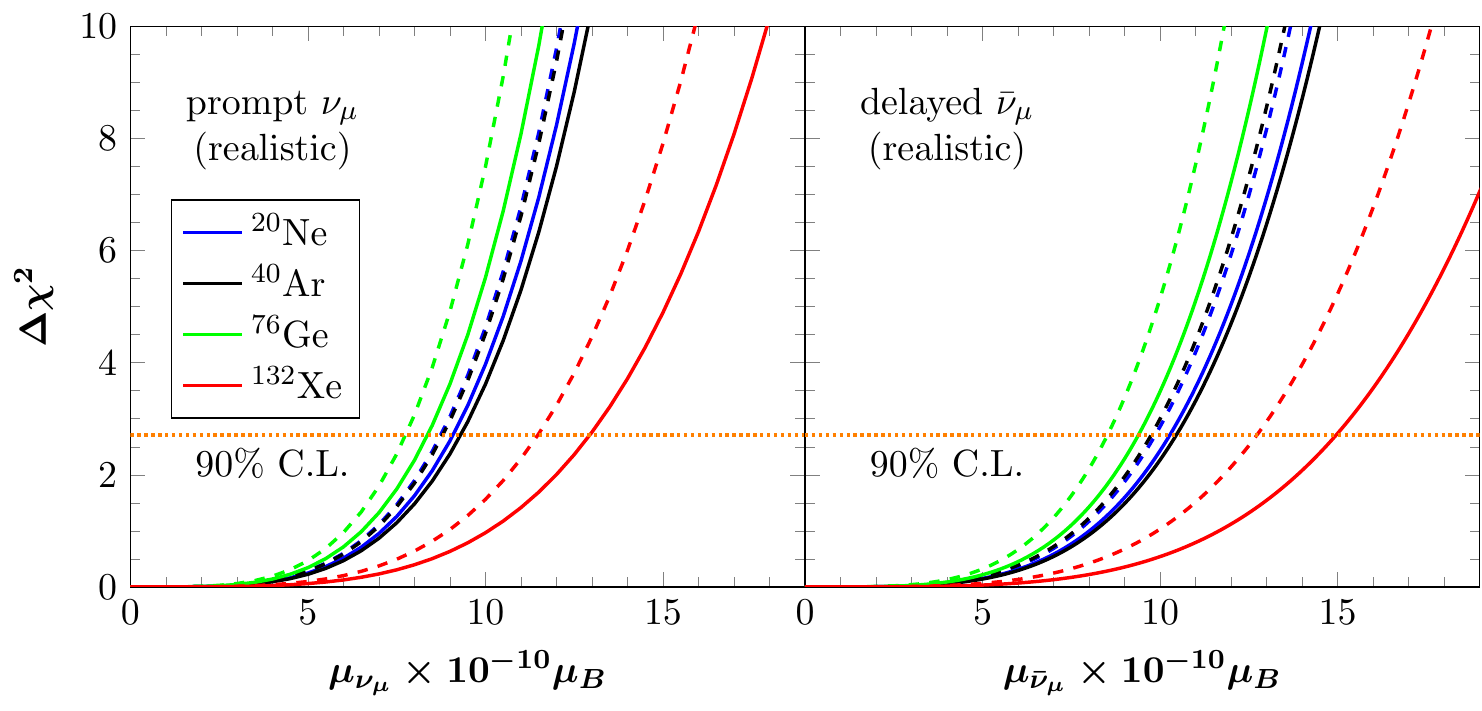}
\includegraphics[width=\textwidth]{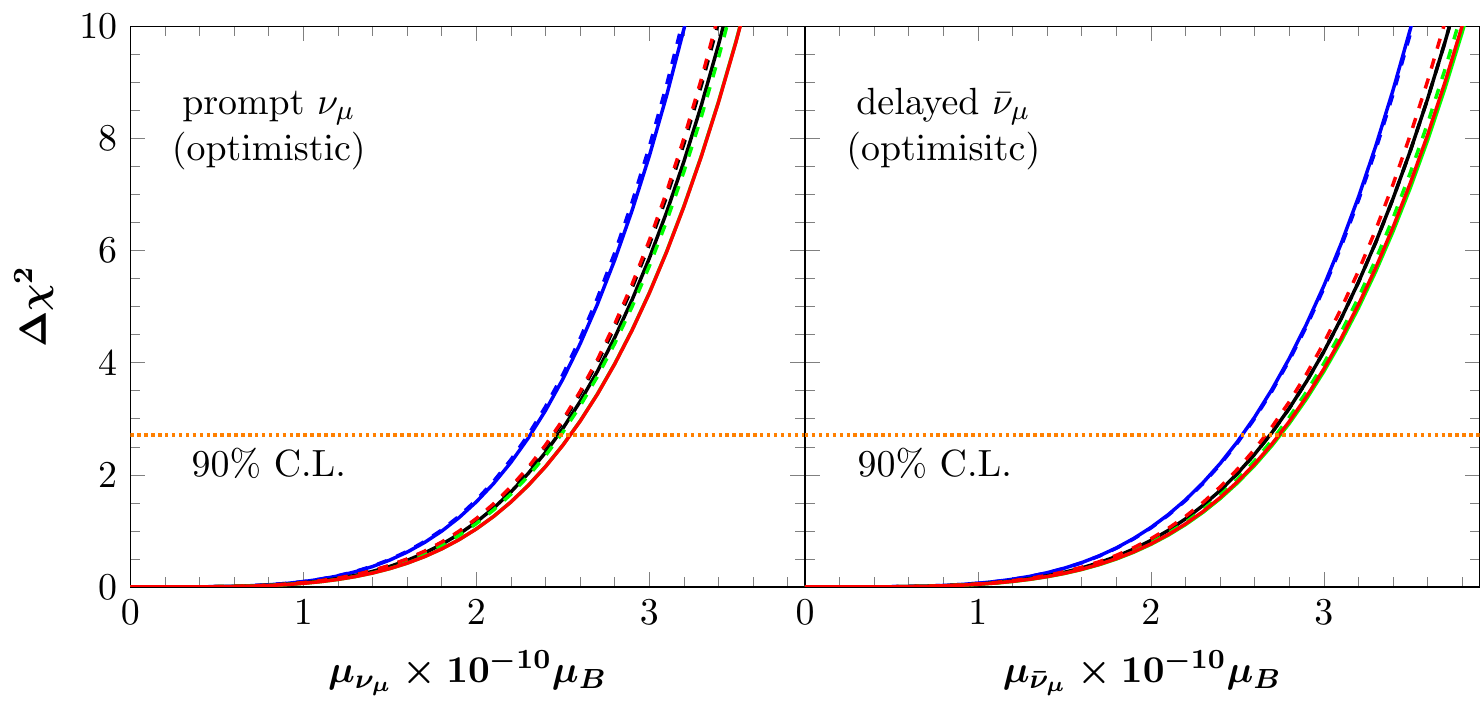}
\caption{$\Delta \chi^2$ profiles for a neutrino magnetic moment,
  $\mu_{\nu_{\mu}}$ in units of $10^{-10} \mu_B$, of the COHERENT
  experiment, assuming various nuclear detectors. The same conventions as
  in Fig.  \ref{fig.sw} are used. }
\label{fig.magn}
\end{figure}

\begin{figure}[ht!]
\centering
\includegraphics[width=\textwidth]{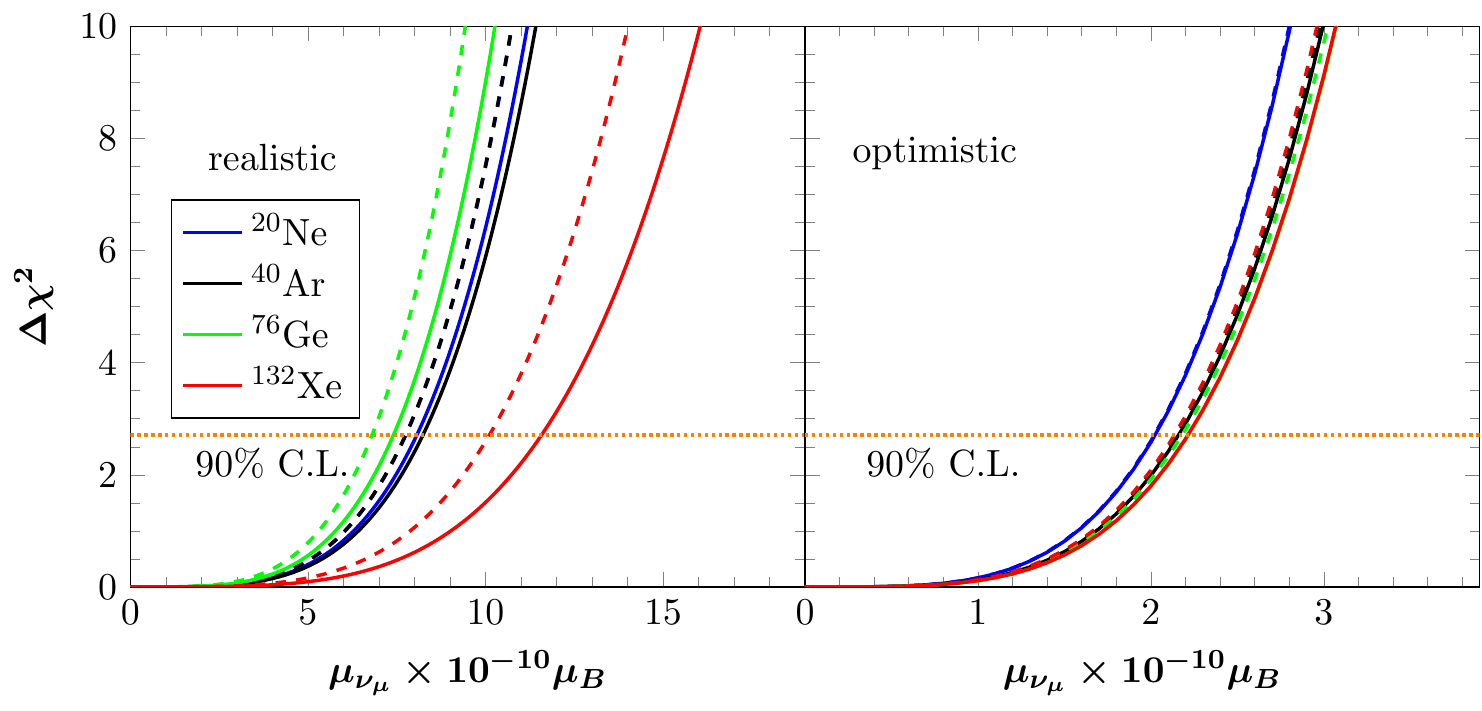}
\caption{ $\Delta\chi^2$ profile for a neutrino magnetic moment,
  $\mu_{\nu_{\mu}}$, in units of $10^{-10} \mu_B$, from the combined
  measurement of the prompt and delayed beams ($\nu_\mu +
  \bar{\nu}_{\mu}$). The same conventions as in Fig.  \ref{fig.sw} are used.  }
\label{fig.magn2}
\end{figure}

The sensitivity is evaluated by assuming that a given experiment
searching for CENNS events, will measure exactly the SM expectation;
thus any deviation is understood as a signature of new
physics. Following~\cite{Tortola:2004vh} we define the $\chi^2$
function as
\begin{equation}
  \chi^2 = \left( \frac{N_{SM}^{events} - N_{\text{tot}}^{events}(\mu_{\nu_{\alpha}})}{\delta N^{events}_{SM}} \right)^2\, .
\end{equation}
By employing the aforementioned method, we find that the COHERENT
experiment could provide useful complementary limits on
$\mu_{\nu_{\mu}}$.  On the other hand, the sensitivity to
$\mu_{\nu_{e}}$, is not expected to be as good as that of reactor
experiments~\cite{Wong:2006nx,Beda:2012zz}.  However, a combined
analysis of the prompt and delayed muon-neutrino beams ($\nu_\mu +
\bar{\nu}_\mu$), could help to further improve the sensitivity to a
neutrino magnetic moment.  The same applies to the combination of
different detectors using the same neutrino source.  For different
nuclear targets, the present results are shown in Figs.~\ref{fig.magn}
and~\ref{fig.magn2} and the sensitivities on neutrino magnetic moments
at 90\% C.L. are summarised in Table~\ref{table.magn}.

%
\begin{table*}[t]
\centering
\caption{Upper limits on the neutrino magnetic moment (in units of
  $10^{-10} \mu_B$) at 90\% C.L. expected at the COHERENT experiment
  for the realistic (optimistic) case. The results indicated 
 with ($^{\rm comb}$) are obtained from a combined measurement of the 
 prompt and delayed beams.}
\begin{tabular}{{c|cccc}}
\hline \hline
Nucleus & $^{20}\mathrm{Ne}$ &  $^{40}\mathrm{Ar}$ & $^{76}\mathrm{Ge}$ & $^{132}\mathrm{Xe}$\\
\hline
$ \mu_{\nu_\mu}$          &  ~9.09 [2.31] &  ~9.30 [2.47] & ~8.37 [2.54] & 12.94 [2.54] \\
$ \mu_{\bar{\nu}_\mu}$  & 10.28 [2.53] & 10.46 [2.69] & ~9.39 [2.75] & 14.96 [2.74] \\
$ \mu_{\nu_e}$            & 10.22 [2.44] & 10.55 [2.60] & ~9.46 [2.68] & 15.20 [2.68]  \\
 \hline 
$ \mu_{\nu_\mu}^{\rm comb} $ & ~8.07 [2.02] & ~8.24 [2.16] & ~7.41 [2.22] & 11.58 [2.21] \\
\hline \hline
\end{tabular}
\label{table.magn}
\end{table*}
%
%
%
%

The sensitivity to neutrino magnetic moments has also been computed
for the case of a combined measurement with different target
nuclei. In this framework, we take advantage of the multitarget
strategy of the COHERENT
experiment~\cite{Bolozdynya:2012xv,Akimov:2013yow} and define the
$\chi^2$ as
\begin{equation}
 \chi^2 = \sum_{nuclei} \left( \frac{N_{SM}^{events} - N_{\text{tot}}^{events}(\mu_{\nu_{\alpha}})}{\delta N_{SM}^{events}} \right)^2\,.
\end{equation}
Assuming two nuclear targets at a time and taking into consideration
the experimental technologies discussed previously, we have found that
among all possible combinations the most stringent sensitivity
corresponds to a combined measurement of $^{20}$Ne+$^{76}$Ge, which for
the realistic (optimistic) case reads
\begin{equation}
\mu_{\nu_\mu}= 6.48 \, (1.77) \times 10^{-10} \mu_B \quad 90\% \, \text{C.L.} 
\end{equation} 
The above sensitivity is better than the case with only one detector.
Notice also that the optimistic sensitivity shown here gives an idea
to the potential constraint that could be achieved by improving the
experimental setup. Moreover, a combined measurement of all possible target nuclei would
lead to somewhat better expected sensitivities, i.e.,
\begin{equation}
\mu_{\nu_\mu}= 5.87 \, (1.52) \times 10^{-10} \mu_B \quad  90\% \, \text{C.L.}
\end{equation}

\begin{figure}[t]
\centering
\includegraphics[width=0.47\textwidth]{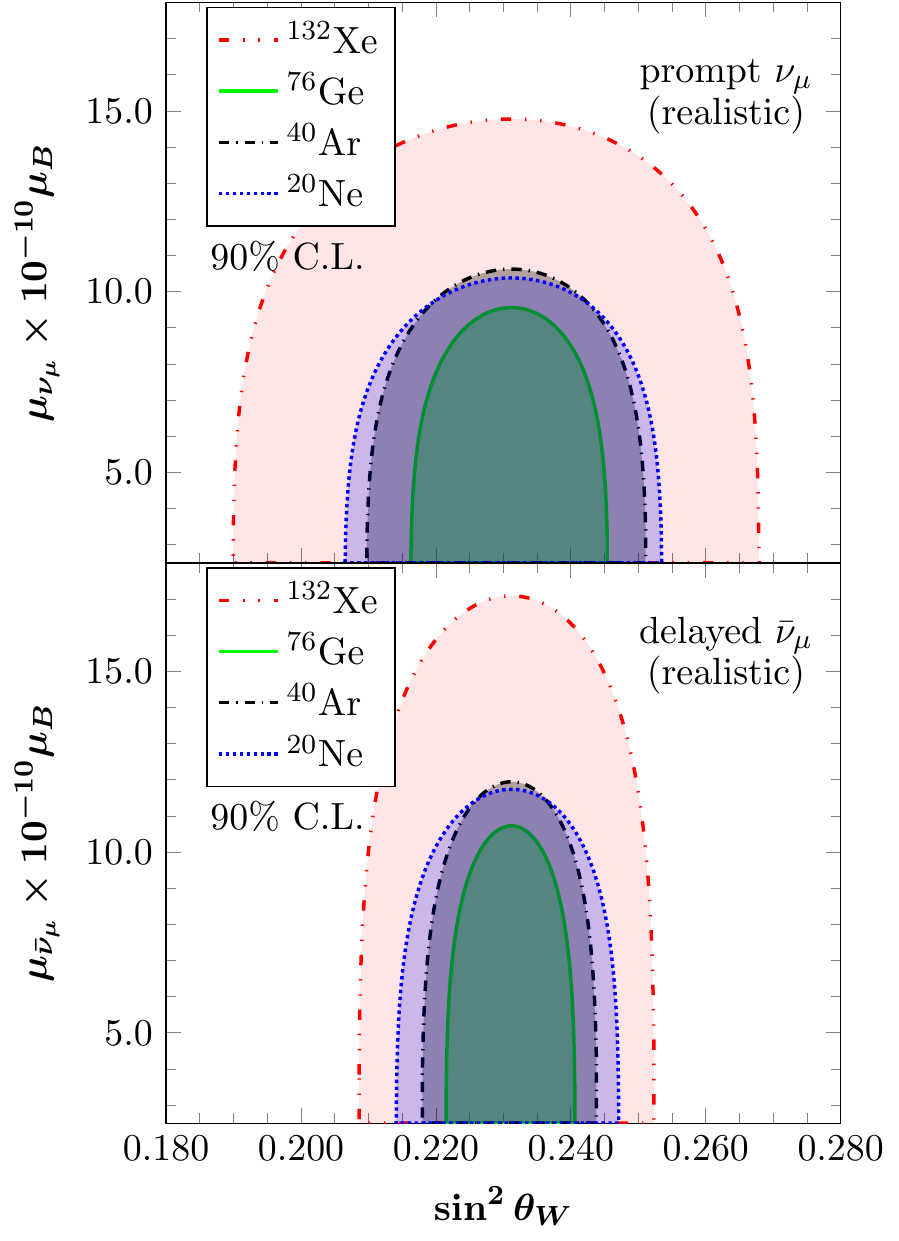}
\includegraphics[width=0.47\textwidth]{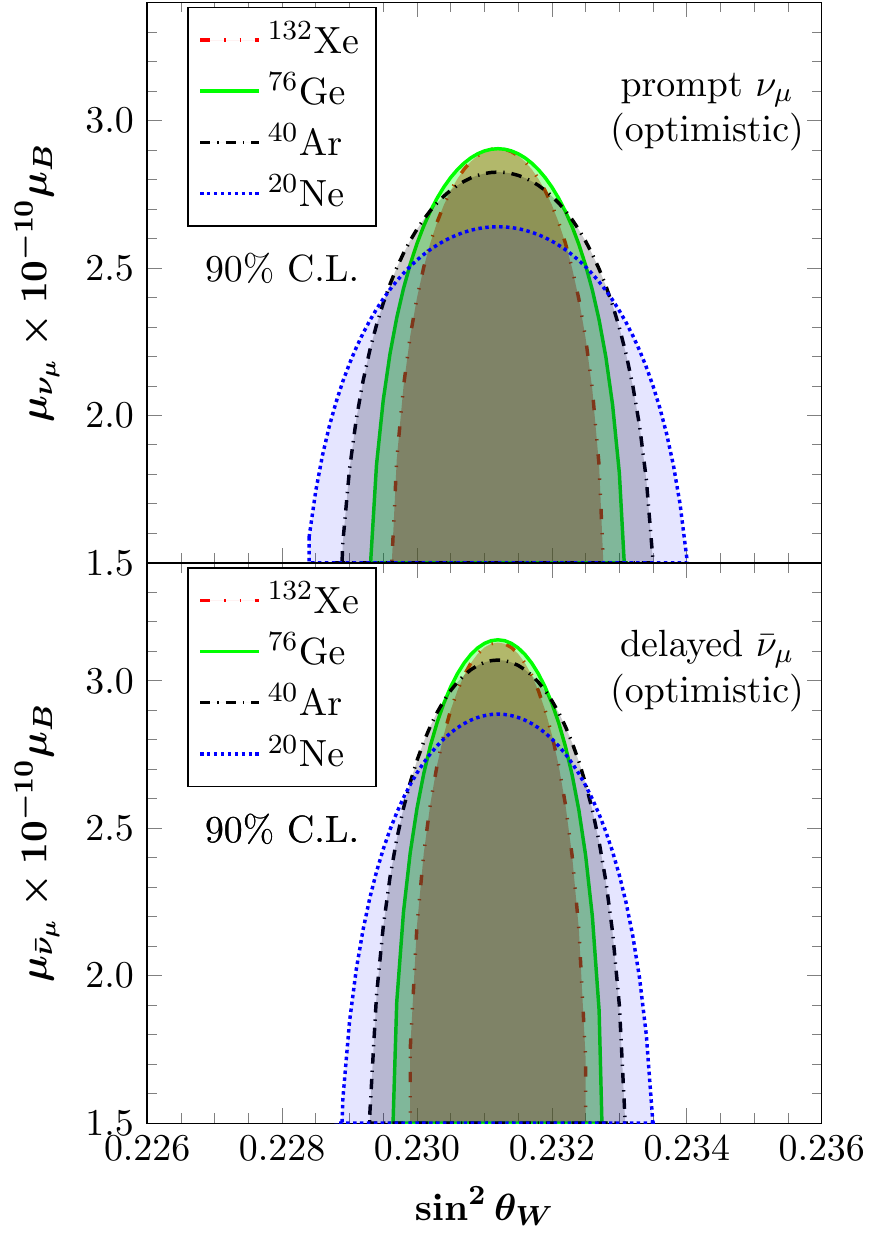}
\caption{The $\mu_{\nu_{\mu}}(\mu_{\bar{\nu}_{\mu}})$-$\sin^2
  \theta_W$ contours obtained from a two parameter $\chi^2$
  analysis. Allowed regions are shown for $90\%$ C.L. Left (right)
  panels account for the realistic (optimistic) case, while the upper
  (lower) panels refer to the prompt (delayed) flux.}
\label{fig.6}
\end{figure}
%

\begin{figure}[t]
\centering
\includegraphics[width=0.47\textwidth]{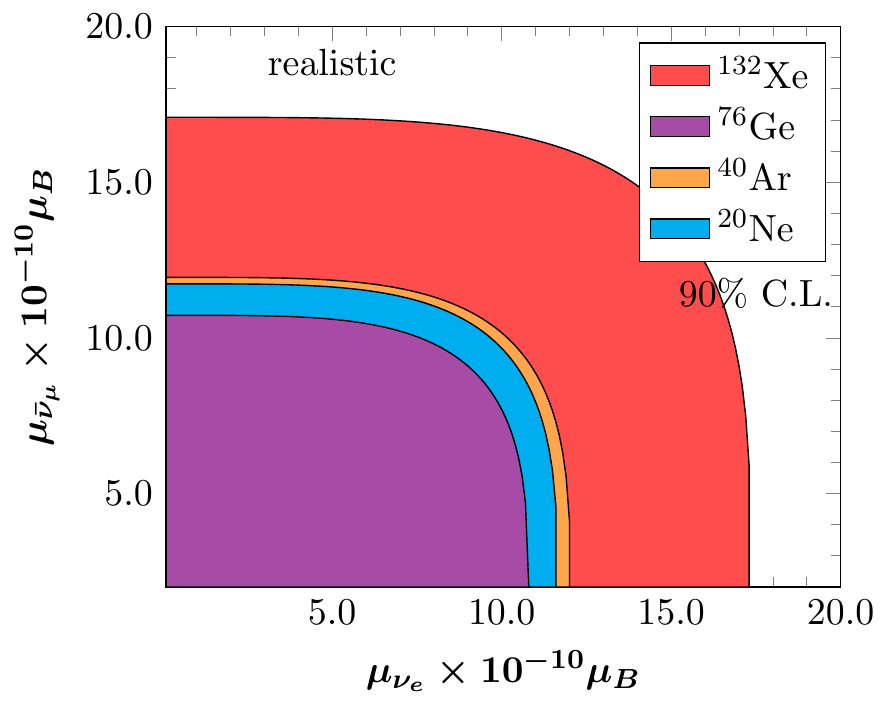}
\includegraphics[width=0.47\textwidth]{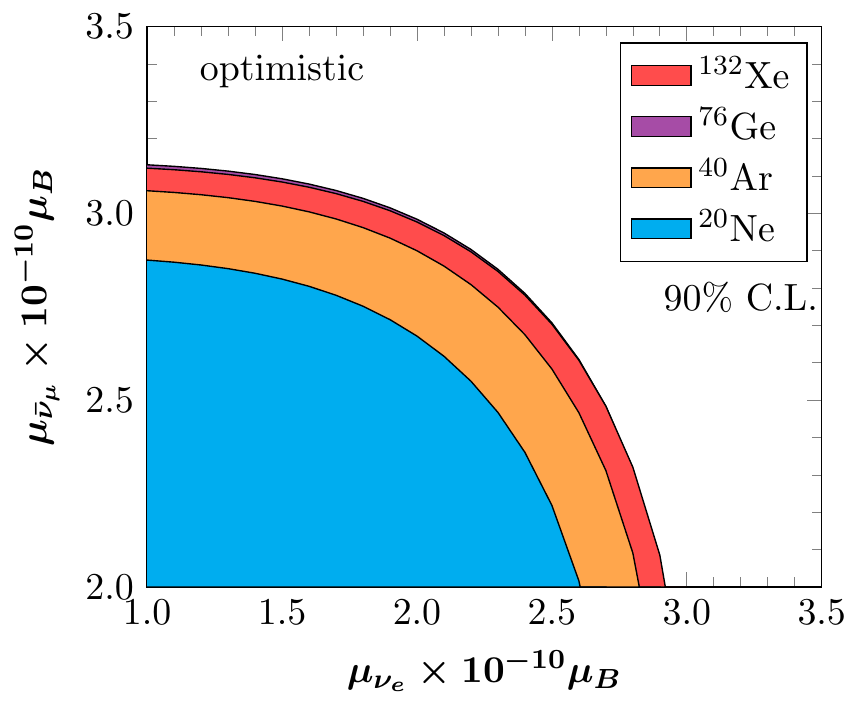}
\caption{The $\mu_{\bar{\nu}_{\mu}}$-$\mu_{\nu_{e}}$
  contours obtained from a two parameter $\chi^2$ analysis. Allowed
  regions are shown for 90\% C.L. Left (right) panel accounts for the
  realistic (optimistic) case.}
\label{fig.7}
\end{figure}

Eventually, we explore the possibility of varying more than one parameter at
the same time.  To this aim, a $\chi^2$ analysis is performed, but in
this case the fitted parameters were simultaneously varied. Within
this context, the contours of the $\sin^2 \theta_W-\mu_\nu$ parameter
space at 90\% C.L. are illustrated in Fig.~\ref{fig.6}. Finally, in
Fig.~\ref{fig.7} the allowed regions of the parameter space in the
$\mu_{\bar{\nu}_{\mu}}$--$\mu_{\nu_{e}}$ plane are shown, where the
corresponding results have been evaluated at 90\% C.L.

%

\section{Discussion and Conclusions}

We have studied the sensitivities on Majorana neutrino magnetic
moments attainable through neutral current coherent neutrino-nucleus
scattering cross section calculations at the Spallation Neutron Source.
Regarding the meaning of the parameter ${\mu_{eff}}$ describing the
effective neutrino magnetic moment, in general this
can be expressed through neutrino amplitudes of positive and negative
helicity states (which we denote as the $3-$vectors $a_{+}$ and
$a_{-}$, respectively) and the magnetic moment matrix,
$\lambda$. Within this notation, the effective neutrino magnetic
moment reads~\cite{Grimus:2002vb}
\begin{equation}
\mu_{eff}^2 = a^\dagger_{+}\lambda\lambda^\dagger a_{+} + 
            a^\dagger_{-}\lambda\lambda^\dagger a_{-}\, . 
\end{equation}
In this work, we mainly focus on the muon-neutrino signal. Then,
considering the muon-neutrino as having a Majorana nature, we have, in
the flavor basis
\begin{equation}
\mu_{eff}^2 = |\Lambda_{e}|^2 + |\Lambda_{\tau}|^2 \, , 
\end{equation}
with $|\Lambda_{e}|$ and $|\Lambda_{\tau}|$ being the elements of the
neutrino transition magnetic moment matrix $\lambda$ describing the
corresponding transitions from the muon-neutrino to the tau and
electron antineutrino states, respectively. The latter expression can be
translated into the mass basis through a rotation by using the
leptonic mixing matrix. An analogous expression can be found for
purely electron neutrino states, for example. One can see that the
limits on the effective neutrino magnetic moment obtained from
neutrino experiments are in reality a restriction on a combination of
physical observables. In this sense, an improvement in the muon
effective neutrino magnetic moment will contribute towards improving
the constraints on the physical observables through a combined
analysis of neutrino data.  A full description of this formalism can
be found in Ref.~\cite{Grimus:2002vb}.

The sensitivities we have extracted are obtained by means of a simple
$\chi^2$ analysis employing realistic nuclear structure calculations
within the QRPA, for the
evaluation of the coherent cross section.  We find that current limits
on the muon-neutrino magnetic moment, $\mu_{\nu_\mu}$, can be improved
by half an order of magnitude.  In addition, we show that the SNS
allows for a competitive determination of the electroweak mixing angle
$\theta_W$. Moreover, the COHERENT proposal may provide an excellent
probe for investigating other electromagnetic neutrino properties,
such as the neutrino charge radius (see the appendix).  In view of the
operation of proposed sensitive neutrino experiments (e.g. COHERENT)
our results, presented for various choices of experimental setups and
target materials, may contribute towards a deeper understanding of
so-far hidden neutrino properties.

\acknowledgments This work was supported by Spanish MINECO under
Grants FPA2014-58183-P and MULTIDARK CSD2009-00064 (Consolider-Ingenio
2010 Programme), EPLANET, the CONACyT Grant No. 166639 (Mexico), and
the Generalitat Valenciana Grant No. PROMETEOII/2014/084. D. K. P. was
supported by \textit{"Prometeu per a grups d' investigaci\'o d'
  Excel$\cdot$l\`{e}ncia de la Conseller\'{i}a d' Educaci\'o, Cultura
  i Esport, CPI-14-289"} (Grant No. GVPROMETEOII2014-084).  DKP is
grateful to Dr. Athanasios Hatzikoutelis for stimulating dis-
cussions. M. T. is also supported by a Ramon y Cajal contract of the
Spanish MINECO.


\section*{APPENDIX Sensitivity to the Neutrino Charge Radius}
\label{Appendix}

Apart from the neutrino magnetic moment, the neutrino charge radius is
another interesting electromagnetic property to be considered.  In
general, the electric form factor allows us to extract nontrivial
information concerning the neutrino electric properties, despite its
neutral electric charge~\cite{Broggini:2012df}. In fact, the
gauge-invariant definition of the neutrino effective charge radius
$\langle r^2_{\nu_{\alpha}}\rangle$, $\alpha=e,\mu, \tau$, was
proposed long ago~\cite{Bardeen:1972vi,Lee:1973fw} as a physical
observable related to the vector and axial vector form factors
involving the EM interaction of a Dirac
neutrino~\cite{Shrock:1982sc,Vogel:1989iv}. In particular, at the
one-loop approximation a correction of a few percent to the weak mixing
angle has been
obtained~\cite{Hirsch:2002uv,bernabeu:2000hf,Barranco:2007ea},

\begin{equation}
\sin^2 \theta_W \rightarrow \sin^2 \overline{\theta_W} + \frac{\sqrt{2} \pi \alpha_{\rm em}}{3 G_F} \langle r^2_{\nu_{\alpha}}\rangle\, .
\end{equation}

\begin{figure}[H]
\centering
\includegraphics[width=\textwidth]{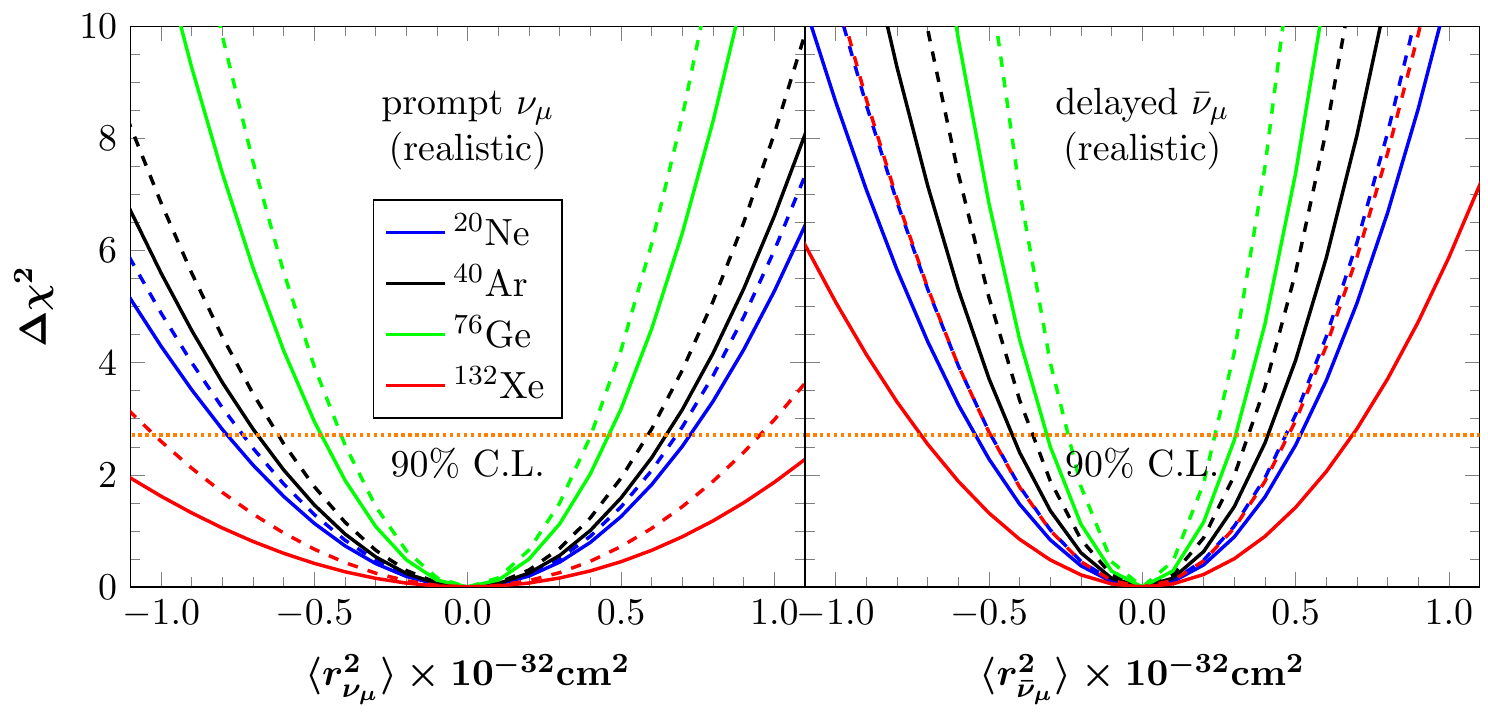}
\includegraphics[width=\textwidth]{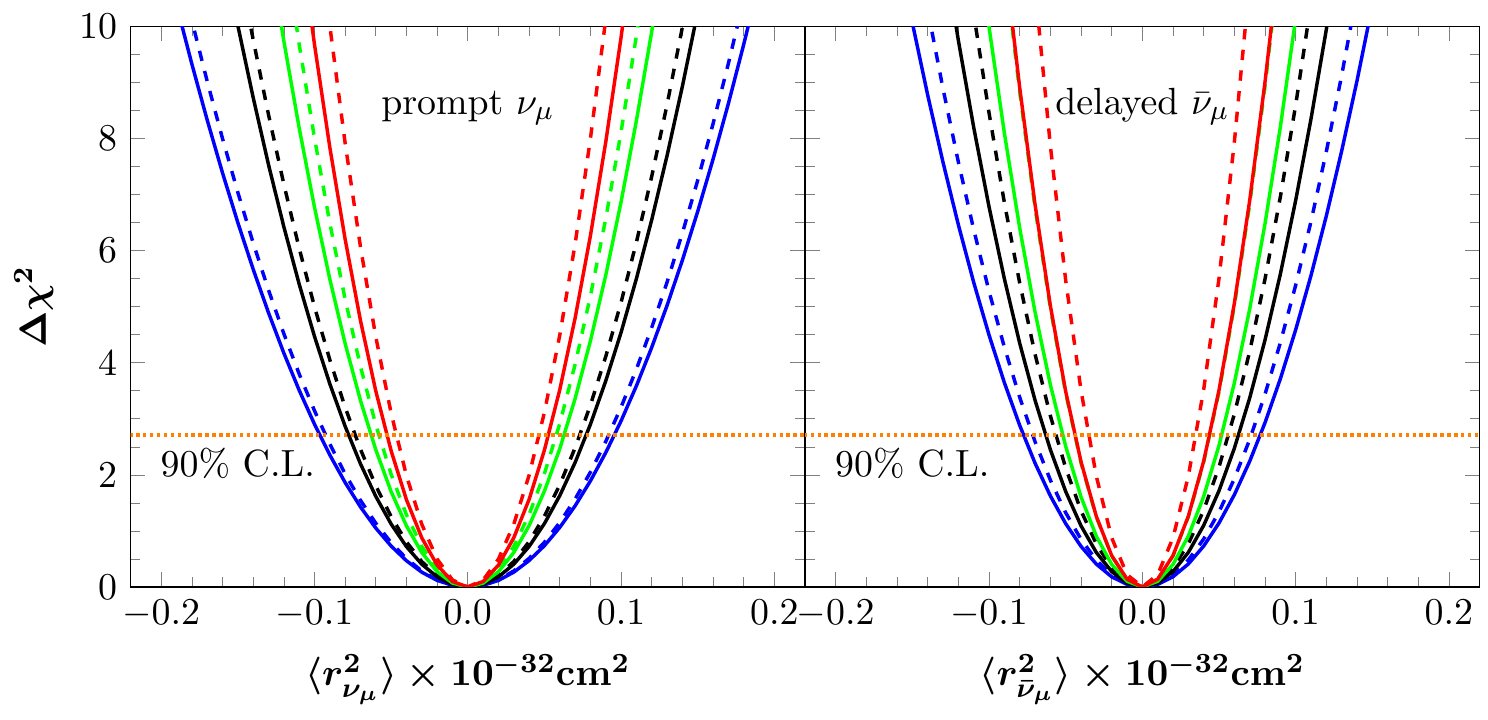}
\caption{$\Delta \chi^2$ profiles for a neutrino charge radius,
  $\langle r_{\nu_\mu}^2 \rangle$ in units of $10^{-32}
  \mathrm{cm^2}$, of the COHERENT experiment, assuming various nuclear
  detectors. Same conventions as in Fig.  \ref{fig.sw} are used. }
\label{fig.rv}
\end{figure}

\begin{figure}[H]
\centering
\includegraphics[width=\textwidth]{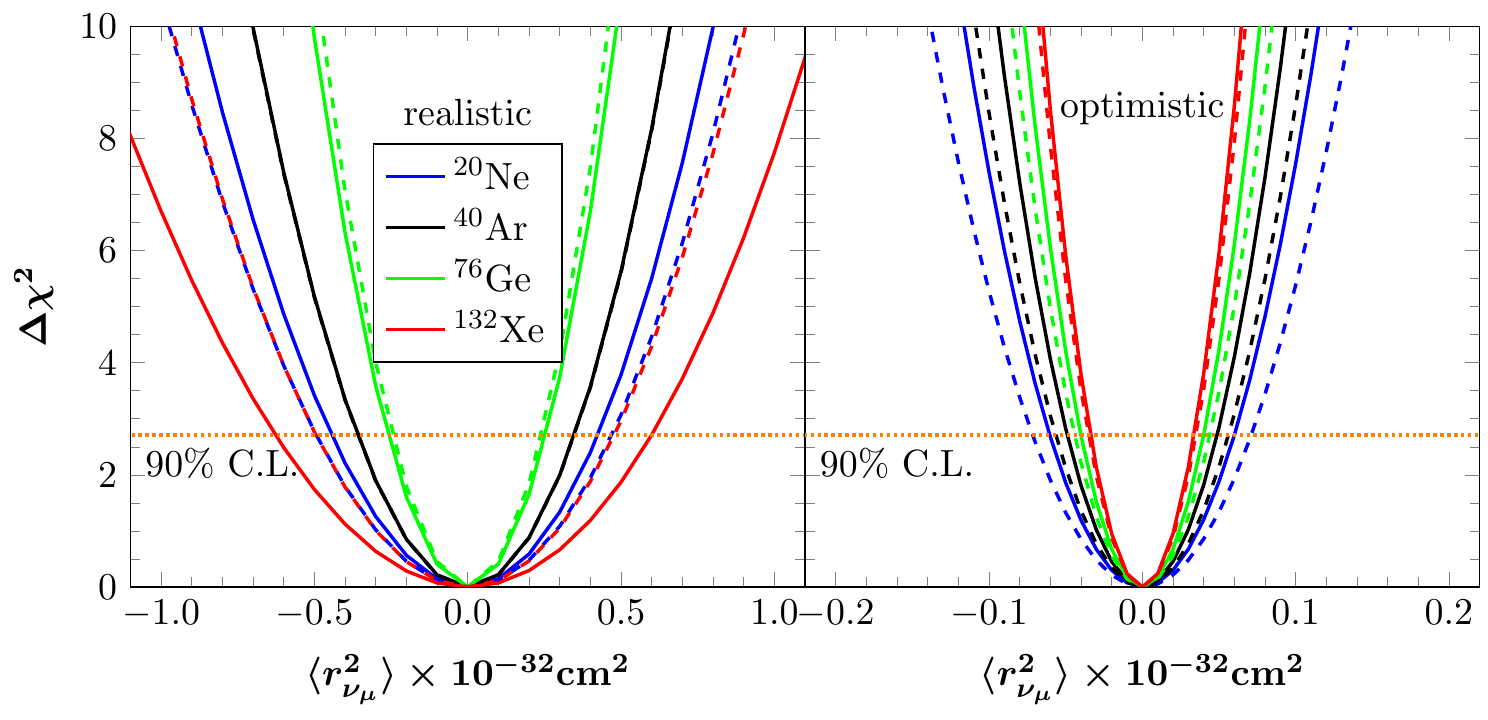}
\caption{ $\Delta\chi^2$ profile for  the neutrino charge radius,
  $\langle r_{\nu_\mu}^2 \rangle$ in units of $10^{-32} \mathrm{cm^2}$,
from the combined measurement of the prompt and delayed beams ($\nu_\mu + \bar{\nu}_{\mu}$). 
Same conventions as in Fig. \ref{fig.sw} are used.}
\label{fig.rv2}
\end{figure}

Through CENNS, we estimate for the first time the sensitivity of a low
energy SNS experiment to constrain the neutrino charge radius. The
obtained bounds are derived in the context of a $\chi^2$ analysis in
the same spirit of the discussion made above and they are presented in
Figs.~\ref{fig.rv} and~\ref{fig.rv2} and listed in
Table~\ref{table.rv2}.  As expected, the results behave similarly to
the case of the weak mixing angle; thus we conclude that for the
realistic (optimistic) case a 100 kg $^{76}$Ge (heavy $^{132}$Xe)
detector at 20 m is required to constrain more significantly the
neutrino charge radius. Furthermore, through a combined measurement of
the prompt and delayed beams ($\nu_\mu + \bar{\nu}_\mu$) an
appreciably improved sensitivity can be reached for $\langle
r^2_{\nu_\mu} \rangle $ in comparison to $\langle r^2_{\nu_e}
\rangle$. These sensitivities are better than current ones (see
Ref.~\cite{Broggini:2012df} and references therein) and depending on
the detector setup may improve by one order of magnitude.
%
%
%
\begin{table*}[t]
\centering
\caption{Expected sensitivities on the neutrino charge radius (in
  units of $10^{-32} \mathrm{cm^2}$) from the analysis of the COHERENT
  experiment. The limits are presented at 90\% C.L. for the realistic
  (optimistic) case. The results indicated with
    ($^{\rm comb}$) are obtained from a combined measurement of the
    prompt and delayed beams. }
\noindent
\begin{tabular}{{c|cccc}}
\hline \hline
Nucleus & $^{20}\mathrm{Ne}$ &  $^{40}\mathrm{Ar}$ & $^{76}\mathrm{Ge}$ & $^{132}\mathrm{Xe}$\\
\hline
\multirow{2}{*}{$ \langle r^2_{\bar{\nu}_\mu} \rangle $} & -0.55 -- 0.52 & -0.43 -- 0.41  & -0.31 -- 0.30 & -0.72 -- 0.69\\
& [-0.08 -- 0.08] & [-0.06 -- 0.06] & [-0.05 -- 0.05] & [-0.04 -- 0.04]\\

\multirow{2}{*}{$ \langle r^2_{\nu_\mu} \rangle $} & -0.79 -- 0.73 & -0.69 -- 0.65 & -0.48 -- 0.46 & -1.31 -- 1.20\\
&  [-0.10 -- 0.10] & [-0.08 -- 0.08] & [-0.06 -- 0.06] & [-0.05 -- 0.05]\\

\multirow{2}{*}{$\langle r^2_{\nu_e} \rangle $}      & -0.65 -- 0.61 & -0.53 -- 0.50  & -0.38 -- 0.37 & -0.90 -- 0.85 \\
 & [-0.09 -- 0.09] & [-0.07 -- 0.07] & [-0.06 -- 0.06] & [-0.05 -- 0.05]\\
\hline
\multirow{2}{*}{ $\langle r^2_{\nu_\mu} \rangle^{\rm comb} $}   & -0.44 -- 0.42 &  -0.36 -- 0.35  & -0.26 -- 0.26 &  -0.63 -- 0.60 \\
& [-0.06 -- 0.06] & [-0.05 -- 0.05] & [-0.04 -- 0.04] & [-0.03 -- 0.03]\\
 \hline \hline
\end{tabular}
\label{table.rv2}
\end{table*}
%
%
%
%
 \bibliographystyle{JHEP}   

\end{document}